# LIFSHITZ TAILS FOR SPECTRA OF ERDŐS–RÉNYI RANDOM GRAPHS

### By Oleksiy Khorunzhiy, Werner Kirsch and Peter Müller[1]

*Université de Versailles, Universität Bochum and University of California, Irvine*

We consider the discrete Laplace operator $\Delta^{(N)}$ on Erdős–Rényi random graphs with $N$ vertices and edge probability $p/N$. We are interested in the limiting spectral properties of $\Delta^{(N)}$ as $N \to \infty$ in the subcritical regime $0 < p < 1$ where no giant cluster emerges. We prove that in this limit the expectation value of the integrated density of states of $\Delta^{(N)}$ exhibits a Lifshitz-tail behavior at the lower spectral edge $E = 0$.

**1. Introduction.** The last decades have seen a growing interest in spectral properties of linear operators defined on graphs, mostly of the adjacency matrix or the graph Laplacian [10, 12, 13, 29]. The aim is to see how properties of the graphs are reflected in properties of the operators and vice versa.

Spectral properties of random graphs, however, still remain to be uncovered to a large extent. The mostly recent works [1, 3, 9, 23, 34, 35] deal with random subgraphs of an infinite graph, such as obtained by a percolation model. Their results range from ergodic properties of the spectrum to the existence and regularity properties of the integrated density of states, as well as its asymptotic behavior near spectral edges.

A different prototype of random graphs was introduced by Erdős and Rényi [15], see also [4, 16] for more recent accounts. Here, one is interested in a scaling limit $N \to \infty$ of an ensemble of graphs with $N$ labeled vertices and an $N$-dependent probability measure. The problem is to get spectral information on the Laplacian or other matrices associated with the graph

Received April 2005.
[1]Supported in part by the Deutsche Forschungsgemeinschaft (DFG) Grant Mu 1056/2–1.

in this limit; see, for example, [2, 17, 20, 24]. As compared to the situation described in the previous paragraph, this one here shares more similarities to the spectral theory of large random matrices, which was originated by Wigner [36, 37]. In these studies the primary questions are related to the existence and explicit form of the mean eigenvalue distribution function of $N \times N$-random matrices in the limit $N \to \infty$ or, in other terms, of the limiting integrated density of states.

In the present paper we study a problem that joins the two branches described. We consider the discrete Laplace operator (the graph Laplacian) on Erdős–Rényi random graphs and show in Theorem 2.5 that the asymptotic behavior of its limiting integrated density of states at the lower spectral edge is given by a Lifshitz tail with Lifshitz exponent $1/2$. This means that the occurrence of eigenvalues right above the lower spectral edge is a large-deviation event. The dominant contribution to the Lifshitz tail is provided by the linear clusters of Erdős–Rényi random graphs. We refer to Remark 2.6(i) and (ii) for a brief outline of our proof. Such a strong probabilistic suppression of eigenvalues near a spectral edge was first quantitatively described by I. M. Lifshitz in the physics literature to account for certain electronic properties in disordered materials [26, 27]. It is nowadays a well-understood phenomenon in the mathematical theory of random Schrödinger operators [8, 22, 25, 30, 33].

**2. Model and result.** Given a natural number $N \geq 2$ and a positive real $p \in \, ]0, N[$, we consider *Erdős–Rényi random graphs* $\mathcal{G}^{(N)}$ with $N$ vertices and edge probability $p/N$. There are many interesting phenomena when allowing $p$ to grow with $N$, see, for example, [4, 15], but in this paper we consider the *sparse* case where $p$ is fixed and does not depend on $N$. The graph $\mathcal{G}^{(N)}$ is a random subgraph of the complete graph $\mathcal{K}^{(N)}$ with $N$ labeled vertices. Edges are distributed independently in $\mathcal{G}^{(N)}$ with the same probability $p/N$. In other words, if $\mathcal{S}_M^{(N)}$ is any given subgraph of $\mathcal{K}^{(N)}$ with $M$ edges, then it is realized by $\mathcal{G}^{(N)}$ with probability

$$(2.1) \qquad \mathbb{P}_p^{(N)}\{\mathcal{G}^{(N)} = \mathcal{S}_M^{(N)}\} = \left(\frac{p}{N}\right)^M \left(1 - \frac{p}{N}\right)^{\binom{N}{2} - M}.$$

The parameter range $]1, \infty[$ for $p$ is called the *supercritical regime*, where there is an emerging giant cluster as $N \to \infty$ [4, 15]. Here, we say that a subgraph of $\mathcal{G}^{(N)}$ is a *cluster*, if it is a maximally connected subgraph of $\mathcal{G}^{(N)}$. By convention, we want to include isolated vertices as one-vertex clusters in this notion, too. In contrast to the supercritical regime, the *subcritical regime* $p \in \, ]0, 1[$ has the property that the fraction of vertices which are either isolated or belong to tree clusters tends to one as $N \to \infty$ [4, 15].

Given any two different vertices $i, j \in \{1, \ldots, N\}$ of $\mathcal{K}^{(N)}$, $i \neq j$, let us denote the edge connecting $i$ and $j$ by the unordered pair $[i,j]$. We write $e_{[i,j]}^{(N)}$ for the random variable which is one, if the edge $[i,j]$ is present in $\mathcal{G}^{(N)}$. Otherwise, $e_{[i,j]}^{(N)}$ is zero. Hence, the $e^{(N)}$'s are i.i.d. Bernoulli random variables with parameter $p/N$ under the measure $\mathbb{P}_p^{(N)}$.

DEFINITION 2.1. The *graph Laplacian* $\Delta^{(N)} \equiv \Delta(\mathcal{G}^{(N)})$ of Erdős–Rényi random graphs $\mathcal{G}^{(N)}$ is the random linear operator on $\mathbb{C}^N$ with matrix elements

$$(2.2) \qquad \Delta_{ij}^{(N)} = \left( \sum_{l=1, l \neq i}^{N} e_{[i,l]}^{(N)} \right) \delta_{ij} - e_{[i,j]}^{(N)} (1 - \delta_{ij})$$

for all $i, j \in \{1, \ldots, N\}$ in the canonical basis of $\mathbb{C}^N$. Here $\delta_{ij} = 1$ if $i = j$ and zero otherwise denotes the Kronecker delta.

REMARK 2.2. (i) The diagonal matrix elements of $\Delta^{(N)}$ specify the random vertex degrees in $\mathcal{G}^{(N)}$. If we denote the corresponding diagonal matrix by $D^{(N)}$, then (2.2) can be rewritten as

$$(2.3) \qquad \Delta^{(N)} = D^{(N)} - A^{(N)},$$

where $A^{(N)}$ is the adjacency matrix of the graph.

(ii) The Laplacian $\Delta^{(N)}$ is nonnegative, as follows from its quadratic form

$$(2.4) \qquad \langle \varphi, \Delta^{(N)} \varphi \rangle = \tfrac{1}{2} \sum_{\substack{i,j=1 \\ i \neq j}}^{N} e_{[i,j]}^{(N)} |\varphi_i - \varphi_j|^2$$

for all $\varphi \in \mathbb{C}^N$, where $\langle \cdot, \cdot \rangle$ stands for the standard scalar product in $\mathbb{C}^N$.

(iii) We conclude from (ii) that the dimension of the kernel of $\Delta^{(N)}$ is equal to the (random) number of clusters in $\mathcal{G}^{(N)}$—the corresponding eigenvectors are constant within any cluster.

(iv) Being a random self-adjoint and nonnegative $N \times N$-matrix, $\Delta^{(N)}$ possesses $N$ nonnegative eigenvalues $\{\lambda_j^{(N)}\}_{j \in \{1, \ldots, N\}}$ which are, of course, random variables. The normalized eigenvalue counting function

$$(2.5) \qquad \sigma_p^{(N)}(E) := \mathbb{E}_p^{(N)} \big( N^{-1} \# \{ j \in \{1, \ldots, N\} : \lambda_j^{(N)} \leq E \} \big)$$

measures the average fraction of eigenvalues that do not exceed a given $E \in \mathbb{R}$. Here, $\mathbb{E}_p^{(N)}$ denotes the mathematical expectation with respect to the probability measure $\mathbb{P}_p^{(N)}$.

The quantity we are interested in is defined in the following:

LEMMA 2.3. *Given any $p \in\ ]0, \infty[$, there exists a right-continuous distribution function $\sigma_p \colon \mathbb{R} \to [0, 1]$, which is called the* integrated density of states *of the Laplacian for Erdős–Rényi random graphs, such that*

$$(2.6) \qquad \sigma_p(E) = \lim_{N \to \infty} \sigma_p^{(N)}(E)$$

*holds for all $E \in \mathbb{R}$, except for the at most countably many discontinuity points of $\sigma_p$.*

REMARK 2.4. (i) The lemma is proven in Section 5, using the known fact [20] that the moments of $\sigma_p^{(N)}$ converge as $N \to \infty$. An alternative approach to the proof of Lemma 2.3 via a resolvent-generating function was previously suggested in [20].

(ii) Remark 2.2(ii) implies that $\sigma_p(E) = 0$ for all $E < 0$.

In this paper we focus on the subcritical regime $p \in\ ]0, 1[$, where there is no emerging giant cluster as $N \to \infty$. Our main result is stated in the next theorem.

THEOREM 2.5. *Let $p \in\ ]0, 1[$. Then, $\sigma_p$ has a Lifshitz tail at the lower edge of the spectrum, $E = 0$, with a Lifshitz exponent $1/2$, that is,*

$$(2.7) \qquad \lim_{E \downarrow 0} \frac{\ln |\ln[\sigma_p(E) - \sigma_p(0)]|}{\ln E} = -\frac{1}{2}.$$

REMARK 2.6. (i) Theorem 2.5 follows from Lemmas 3.1 and 4.1 below, which provide upper and lower bounds for $\sigma_p$. Their proof is close in spirit to that in [23] for Laplacians on bond-percolation graphs. The bounds of Lemma 3.1 and Lemma 4.1 yield

$$(2.8) \qquad p - 1 - \ln p \le -\lim_{E \downarrow 0} \frac{\ln[\sigma_p(E) - \sigma_p(0)]}{E^{-1/2}} \le 2\sqrt{3}(p - \ln p)$$

for all $E \in\ ]0, \infty[$ and all $p \in\ ]0, 1[$, which is a slightly stronger statement than (2.7).

(ii) The main tool of the proof is a cluster decomposition of the block-diagonal Laplacian. The value $1/2$ for the Lifshitz exponent relates to the fact that the asymptotic behavior of $\sigma_p(E)$ as $E \downarrow 0$ is dominated by the smallest eigenvalues of the linear clusters in Erdős–Rényi random graphs. Indeed, the Cheeger-type lower bound of Lemma A.1 for the smallest nonzero Laplacian eigenvalue of a cluster, which enters the upper bound for $\sigma_p(E) - \sigma_p(0)$, captures the correct size dependence for large linear clusters up to a constant. It also ensures that the smallest nonzero eigenvalue of a linear cluster is among the smallest of nonzero eigenvalues of all clusters of the same size. Together with the exponential decay of the cluster-size distribution, this will yield the desired upper bound. On the other hand, the lower bound is obtained from retaining only the contribution of linear clusters to $\sigma_p(E) - \sigma_p(0)$, which can be estimated in an elementary way.

(iii) Remark 2.2(ii) and (iii) imply that

$$(2.9) \qquad \sigma_p^{(N)}(0) = \frac{\mathbb{E}_p^{(N)}\{\operatorname{Tr} P_0^{(N)}\}}{N}$$

can be interpreted as the mean number density of clusters. Here, we introduced the orthogonal projection $P_0^{(N)}$ on the kernel of $\Delta^{(N)}$ and abbreviated the trace over $N \times N$-matrices by Tr. It is known [4] that the number of clusters grows linearly in $N$ as $N \to \infty$, so $E = 0$ has to be a discontinuity point of $\sigma_p$. The equality

$$(2.10) \qquad \sigma_p(0) = \lim_{N \to \infty} \sigma_p^{(N)}(0)$$

is therefore not guaranteed by Lemma 2.3, but we show in (3.9) below that it is true for every $p \in ]0, 1[$. It seems an open question to us whether (2.10) remains true for $p \in [1, \infty[$—as is the fate of Theorem 2.5 for $p \in [1, \infty[$.

(iv) The papers [2, 20] provide recursion relations for the moments of the integrated-density-of-states measure as $N \to \infty$ for both the adjacency matrix and the Laplacian of Erdős–Rényi random graphs. The asymptotic behavior of the largest eigenvalue of the adjacency matrix was determined in [24], and [17] shows that the integrated density of states has a dense set of discontinuities; see also the numerical results in [2].

(v) More detailed spectral properties of Erdős–Rényi random graphs have been obtained in the theoretical-physics literature. Using the replica trick and other non rigorous arguments, it is argued in [6] that

$$(2.11) \qquad \begin{aligned} &-\lim_{E \downarrow 0} \frac{\ln[\sigma_p(E) - \sigma_p(0)]}{E^{-1/2}} \\ &= -[1 - p(1 - Q_p)]^{1/2} \ln[p(1 - Q_p)] =: g(p) \end{aligned}$$

for all $p \in ]0, \infty[$, where $Q_p$ is the biggest nonnegative solution of the equation $Q = 1 - e^{-pQ}$. If $p \in ]0, 1[$, then $Q_p = 0$ and $g(p)$ simplifies to $-(1-p)^{1/2} \ln p$, which lies in between the bounds provided by (2.8). In [7, 14] the density of states of $\Delta^{(N)}$ was examined by a combination of analytical and numerical methods for general $E > 0$. Their numerical results, however, were not conclusive enough as to deduce the Lifshitz-tail behavior (2.7). The existence of emerging delocalized states in the giant cluster for $p \gg 1$ was addressed in [5, 28].

(vi) Weighted Erdős–Rényi random graphs also occur in some physical applications. The associated graph Laplacian is again given by (2.2), but now the probability distribution $\mathbb{P}_p^{(N)}$ of the edge random variables $e_{[i,j]}^{(N)}$ is more general than Bernoulli. In many cases, it is required to have an atom at zero with weight $1 - p/N$, corresponding to an absent edge, and a finite second moment $\mathbb{P}_p^{(N)}\{(e_{[i,j]}^{(N)})^2\}$ which is of the order $N^{-1}$. Random matrices of this type also fall under the name sparse random matrices. Some of their spectral properties were studied by, for example, [6, 18, 19, 28, 31].

## 3. Estimate from above.

This section serves to establish an upper bound for the integrated density of states. The bound relies on a Cheeger-type inequality and the exponential decay of the cluster-size distribution in the subcritical regime. Both results are included in the Appendix.

LEMMA 3.1.    There exists a constant $C \in {]}0, \infty{[}$ such that, for every $p \in {]}0, 1{[}$, the integrated density of states satisfies the estimate

$$(3.1) \qquad \sigma_p(E) - \sigma_p(0) \leq C \frac{e^{f(p)}}{p} \exp\{-f(p)E^{-1/2}\}$$

for every $E \in {]}0, \infty{[}$, with the strictly positive decay parameter $f(p) := p - 1 - \ln p$.

PROOF.    Fix $p \in {]}0, 1{[}$ and let $E \in {]}0, \infty{[}$ be a continuity point of $\sigma_p$. We introduce the right-continuous Heaviside unit-step function $\Theta$ so that $\Theta(x) = 1$ if $x \geq 0$ and zero otherwise. Appealing to the spectral theorem and the functional calculus, we infer from (2.5) and (2.9) that

$$(3.2) \qquad \begin{aligned} \sigma_p^{(N)}(E) - \sigma_p^{(N)}(0) &= \mathbb{E}_p^{(N)}\{N^{-1}\operatorname{Tr}[\Theta(E - \Delta^{(N)}) - P_0^{(N)}]\} \\ &= \mathbb{E}_p^{(N)}\{[\Theta(E - \Delta^{(N)}) - P_0^{(N)}]_{11}\}. \end{aligned}$$

To get the second equality in (3.2), we evaluated the trace in the canoncial basis and used enumeration invariance of $\mathbb{E}_p^{(N)}$.

The Laplacian $\Delta^{(N)}$ has nonzero off-diagonal matrix elements between vertices in the same cluster only—and so has any function of $\Delta^{(N)}$. Let $\mathcal{C}(1)$ denote the cluster of $\mathcal{G}^{(N)}$ that contains vertex number 1. We define the associated graph Laplacian $\Delta(\mathcal{C}(1))$ as the random linear operator on $\mathbb{C}^N$ whose matrix elements $[\Delta(\mathcal{C}(1))]_{ij}$ coincide with $\Delta_{ij}^{(N)}$ for $i, j \in \mathcal{C}(1)$, but are zero otherwise. Likewise, $P_0(\mathcal{C}(1))$ stands for the orthogonal projector in $\mathbb{C}^N$ on the kernel of $\Delta(\mathcal{C}(1))$. Thus, introducing the characteristic function $\chi_{\Omega_0}$ of the event that vertex number 1 is not isolated, we obtain

$$\sigma_p^{(N)}(E) - \sigma_p^{(N)}(0) = \mathbb{E}_p^{(N)}\{\chi_{\Omega_0}[\Theta(E - \Delta(\mathcal{C}(1))) - P_0(\mathcal{C}(1))]_{11}\}$$

$$(3.3) \qquad \leq \mathbb{E}_p^{(N)}\{\chi_{\Omega_0}\Theta(E - E_{\min}(\mathcal{C}(1)))\}$$

$$\leq \mathbb{E}_p^{(N)}\{\chi_{\Omega_0}\Theta(E - |\mathcal{C}(1)|^{-2})\}.$$

The first inequality in (3.3) follows from the spectral theorem with $E_{\min}(\mathcal{C}(1))$ denoting the smallest nonzero eigenvalue of $\Delta(\mathcal{C}(1))$. The second inequality in (3.3) uses the Cheeger-type estimate of Lemma A.1 in the Appendix, and $|\mathcal{C}(1)|$ counts the number of vertices in the cluster $\mathcal{C}(1)$. The expression in the last line of (3.3) is equal to

$$(3.4) \qquad \sum_{n=m(E)}^{\infty} \mathbb{P}_p^{(N)}\{|\mathcal{C}(1)| = n\} = 1 - \sum_{n=1}^{m(E)-1} \mathbb{P}_p^{(N)}\{|\mathcal{C}(1)| = n\},$$

where $m(E) := \max\{2, \lfloor E^{-1/2} \rfloor\}$ and $\lfloor x \rfloor$ stands for the biggest integer, not exceeding $x \in \mathbb{R}$. Since $E$ was chosen to be a continuity point of $\sigma_p$, we deduce with the help of Lemma 2.3 that

$$(3.5) \qquad \sigma_p(E) - \liminf_{N \to \infty} \sigma_p^{(N)}(0) \leq 1 - \sum_{n=1}^{m(E)-1} n\tau_n(p) = \sum_{n=m(E)}^{\infty} n\tau_n(p).$$

Here we introduced the cluster-size distribution

$$(3.6) \qquad \tau_n(p) := n^{-1} \lim_{N \to \infty} \mathbb{P}_p^{(N)}\{|\mathcal{C}(1)| = n\}$$

of Erdős–Rényi random graphs, whose existence (A.7) and normalization (A.8) is summarized in Lemma A.2 in the Appendix. Taking the limit $E \downarrow 0$ in (3.5) along a sequence of continuity points and appealing to the right-continuity of $\sigma_p$, we infer

$$(3.7) \qquad \liminf_{N \to \infty} \sigma_p^{(N)}(0) \geq \sigma_p(0).$$

On the other hand, the monotonicity and right-continuity of $\sigma_p$ imply, for all $p \in ]0, \infty[$,

$$(3.8) \qquad \limsup_{N \to \infty} \sigma_p^{(N)}(0) \leq \lim_{E \downarrow 0} \limsup_{N \to \infty} \sigma_p^{(N)}(E) = \lim_{E \downarrow 0} \sigma_p(E) = \sigma_p(0),$$

where the limit $E \downarrow 0$ is again taken along a sequence of continuity points. From (3.7) and (3.8), we conclude the existence of the limit

$$(3.9) \qquad \lim_{N \to \infty} \sigma_p^{(N)}(0) = \sigma_p(0)$$

in the subcritical regime $p \in ]0, 1[$. Hence, (3.5) and the exponential decay (A.9) of the cluster-size distribution lead to

$$\sigma_p(E) - \sigma_p(0) \leq \frac{1}{\sqrt{2\pi}p} \sum_{n=m(E)}^{\infty} n^{-3/2} e^{-nf(p)}$$

$$(3.10) \qquad \leq \frac{e^{-m(E)f(p)}}{\sqrt{2\pi p}} \sum_{n=0}^{\infty} [n + m(E)]^{-3/2} e^{-nf(p)}$$

$$\leq \frac{\exp\{-f(p)[E^{-1/2} - 1]\}}{\sqrt{2\pi p}} \sum_{n=2}^{\infty} n^{-3/2}.$$

Finally, the estimate (3.10) extends to all $E \in ]0, \infty[$ by right-continuity. This completes the proof of Lemma 3.1.  □

**4. Estimate from below.** In this section we derive a lower bound for the integrated density of states by retaining only contributions from linear clusters.

LEMMA 4.1.  *Let $p \in ]0, 1[$ and define $F(p) := p - \ln p > 1$. Then the estimate*

$$(4.1) \qquad \sigma_p(E) - \sigma_p(0) \geq \frac{e^{-F(p)}}{2p} \exp\{-2\sqrt{3}F(p)E^{-1/2}\}$$

*holds for every $E \in ]0, \infty[$.*

REMARK 4.2.  The decay parameter $F(p)$ is related to that in Lemma 3.1 by $F(p) = f(p) + 1$.

PROOF OF LEMMA 4.1.  Fix $p \in ]0, 1[$. The right-continuity of $\sigma_p$ and the monotonicity of $\sigma_p^{(N)}$ imply $\sigma_p(E) \geq \limsup_{N \to \infty} \sigma_p^{(N)}(E)$ for *every* $E \in ]0, \infty[$. Together with (3.7) and (3.2), this yields

$$(4.2) \quad \sigma_p(E) - \sigma_p(0) \geq \limsup_{N \to \infty} \mathbb{E}_p^{(N)}\{N^{-1} \mathrm{Tr}[\Theta(E - \Delta^{(N)}) - P_0^{(N)}]\}.$$

The decomposition of the random graph $\mathcal{G}^{(N)}$ into its random clusters $\{\mathcal{C}_k^{(N)}\}_{k \in \{1, \dots, K\}}$ provides us with the relation

$$\mathrm{Tr}[\Theta(E - \Delta^{(N)}) - P_0^{(N)}] = \sum_{k=1}^{K} \mathrm{Tr}[\Theta(E - \Delta(\mathcal{C}_k^{(N)})) - P_0(\mathcal{C}_k^{(N)})]$$

$$(4.3)$$

$$\geq \sum_{k=1}^{K} \Theta(|\mathcal{C}_k^{(N)}| - 2)\Theta(E - E_{\min}(\mathcal{C}_k^{(N)})).$$

The inequality in (4.3) relies on the spectral theorem. Note that one-vertex clusters do not contribute to the right-hand side of the first line in (4.3).

For $n \in \{2, \dots, N\}$, let $\chi_{\mathfrak{L}_n}(\mathcal{C}_k^{(N)})$ be the indicator function of the event that the cluster $\mathcal{C}_k^{(N)}$ is a linear chain with $n \geq 2$ vertices, that is, that it is a connected graph having $n - 2$ vertices with degree 2 and 2 vertices with degree 1. Using (4.2) and (4.3), we then obtain the first inequality of the chain

$$
\begin{aligned}
\sigma_p(E) &- \sigma_p(0) \\
&\geq \limsup_{N \to \infty} \mathbb{E}_p^{(N)} \left\{ N^{-1} \sum_{k=1}^{K} \sum_{n=2}^{\infty} \chi_{\mathfrak{L}_n}(\mathcal{C}_k^{(N)}) \Theta(E - E_{\min}(\mathcal{C}_k^{(N)})) \right\} \\
&\geq \limsup_{N \to \infty} \sum_{n=2}^{\infty} \Theta(E - 12/n^2) \mathbb{E}_p^{(N)} \left\{ N^{-1} \sum_{k=1}^{K} \chi_{\mathfrak{L}_n}(\mathcal{C}_k^{(N)}) \right\} \\
&\geq \frac{1}{M(E)} \limsup_{N \to \infty} \mathbb{P}_p^{(N)} \{ \mathcal{C}(1) \text{ is linear and has } M(E) \text{ vertices} \}.
\end{aligned}
$$

To derive the second inequality in (4.4), we used the upper bound $12/n^2$ for the smallest nonzero Laplacian eigenvalue of a linear chain with $n$ vertices, see, for example, Lemma 2.6(i) in [23]. For the last inequality in (4.4), we introduced $M(E) := \lfloor (12/E)^{1/2} \rfloor + 1$, the smallest integer strictly greater than $(12/E)^{1/2}$, dropped all terms in the $n$-sum, except the one with $n = M(E)$, and observed

$$
\begin{aligned}
\frac{1}{N} \mathbb{E}_p^{(N)} \left\{ \sum_{k=1}^{K} \chi_{\mathfrak{L}_n}(\mathcal{C}_k^{(N)}) \right\} &= \frac{1}{nN} \mathbb{E}_p^{(N)} \left\{ \sum_{j=1}^{N} \chi_{\mathfrak{L}_n}(\mathcal{C}(j)) \right\} \\
&= \frac{1}{n} \mathbb{E}_p^{(N)} \{ \chi_{\mathfrak{L}_n}(\mathcal{C}(1)) \}.
\end{aligned}
$$

Equation (4.5) involves $\mathcal{C}(j)$, the cluster of $\mathcal{G}^{(N)}$ containing vertex $j$, and it exploits enumeration invariance of $\mathbb{P}_p^{(N)}$. Now, elementary combinatorics shows for any $m \in 2, \dots, N$ that

$$
\begin{aligned}
\mathbb{P}_p^{(N)} &\{ \mathcal{C}(1) \text{ is linear and has } m \text{ vertices} \} \\
&= \binom{N-1}{m-1} \frac{m!}{2} \left( \frac{p}{N} \right)^{m-1} \left( 1 - \frac{p}{N} \right)^{(N-3)(m-2)+2(N-2)}.
\end{aligned}
$$

Here the first factor corresponds to the choice of $m - 1$ vertices different from the already fixed vertex number one. The second factor corresponds to ordering these $m$ vertices in a chain. The third factor accounts for the probability to join these $m$ vertices by $m - 1$ edges and the last one assures that there are no other edges joining the $m$ vertices to the remaining $N - m$ vertices. Hence, the limit $N \to \infty$ of (4.6) exists and is given by

$$
\begin{aligned}
\lim_{N \to \infty} \mathbb{P}_p^{(N)} &\{ \mathcal{C}(1) \text{ is linear and has } m \text{ vertices} \} \\
&= \frac{m}{2} p^{m-1} e^{-pm} \lim_{N \to \infty} \frac{(N-1)!}{N^{m-1}(N-m)!} = \frac{m}{2} p^{m-1} e^{-pm}.
\end{aligned}
$$

Inserting this into (4.4), we arrive at

$$(4.8) \qquad \sigma_p(E) - \sigma_p(0) \geq \frac{1}{2p} \exp[-(p - \ln p)M(E)],$$

which implies the lemma. □

## 5. Existence of the integrated density of states.

PROOF OF LEMMA 2.3.   Let $p \in \, ]0, \infty[$. Theorem 2 in [20] establishes the existence and finiteness of the limits

$$(5.1) \quad M_k^\Delta := \lim_{N \to \infty} \int_{[0,\infty[} d\sigma_p^{(N)}(E) E^k = \lim_{N \to \infty} \mathbb{E}_p^{(N)} \{N^{-1} \operatorname{Tr}[(\Delta^{(N)})^k]\},$$

$k \in \mathbb{N}_0$, of all moments of $\sigma_p^{(N)}$ as $N \to \infty$. Being the limit of a sequence of Stieltjes moments, $\{M_k^\Delta\}_{k \in \mathbb{N}_0}$ is itself a sequence of Stieltjes moments associated to some, not necessarily unique, distribution function $\sigma_p$ on $[0, \infty[$. This follows from Theorem 1.1 in [32], see also the statements in Chapter I.2(b) there. We extend $\sigma_p$ to $\mathbb{R}$ by setting it to zero on $]-\infty, 0[$. We will prove the bound

$$(5.2) \qquad M_{2k}^\Delta \leq (c_p k)^{2k}$$

for all $k \in \mathbb{N}$ with some $k$-independent constant $c_p \in \, ]0, \infty[$. This, in turn, guarantees the Carleman condition $\sum_{k=1}^\infty (M_{2k}^\Delta)^{-1/(2k)} = +\infty$ and, by Theorem 1.10 in [32], the uniqueness of the Hamburger (and, hence, the Stieltjes) moment problem. Knowing the uniqueness of $\sigma_p$, the lemma then follows from, for example, Theorem 4.5.5 in [11].

To prove (5.2), we use (2.3) and Hölder's inequality for the Schatten trace norms $\|B\|_q := \{\operatorname{Tr}[(B^*B)^{q/2}]\}^{1/q}$, $q \geq 1$, of complex $N \times N$-matrices, where $B^*$ stands for the adjoint of $B$. For every natural number $N > p$, this yields the bound

$$
\begin{aligned}
(5.3) \qquad \operatorname{Tr}[(\Delta^{(N)})^{2k}] &\leq \sum_{\kappa=0}^{2k} \binom{2k}{\kappa} \|D^{(N)}\|_{2k}^\kappa \|A^{(N)}\|_{2k}^{2k-\kappa} \\
&= (\|D^{(N)}\|_{2k} + \|A^{(N)}\|_{2k})^{2k} \\
&\leq 2^{2k-1} (\|D^{(N)}\|_{2k}^{2k} + \|A^{(N)}\|_{2k}^{2k}).
\end{aligned}
$$

Hence, we get

$$(5.4) \qquad M_{2k}^\Delta \leq 2^{2k-1}(M_{2k}^D + M_{2k}^A),$$

where, thanks to ergodicity,

$$(5.5) \qquad M_{2k}^D := \lim_{N \to \infty} \mathbb{E}_p^{(N)}\{N^{-1} \operatorname{Tr}[(D^{(N)})^{2k}]\} = e^{-p} \sum_{n=0}^\infty \frac{p^n}{n!} n^{2k}$$

is nothing but the $2k$th moment of the Poissonian [4] vertex-degree distribution and $M_{2k}^A := \lim_{N \to \infty} \mathbb{E}_p^{(N)}\{N^{-1} \operatorname{Tr}[(A^{(N)})^{2k}]\}$ is the $2k$th moment of the adjacency matrix. So, (5.2) is implied by (5.4), provided we show

$$(5.6) \qquad \max\{M_{2k}^D, M_{2k}^A\} \le (c_p k/2)^{2k}$$

for all $k \in \mathbb{N}$. Concerning $M_{2k}^D$, this follows from applying the elementary inequality $a^b \le b^b + b^a$ to (5.5). This inequality holds for $a \in ]0, \infty[$ and $b \in [e, \infty[$. Its validity is obvious for $a \le b$, while, for $a > b \ge e$, it can be deduced from Jensen's inequality. The desired bound for $M_{2k}^A$ is established in the proof of Proposition 1 in [17], using a result of [21]; see also [2]. Hence, (5.2) is proven, and so is the lemma. $\quad\square$

## APPENDIX

For completeness and convenience of the reader, we state and prove two auxiliary results in this Appendix, which were needed in the proof of the upper bound in Lemma 3.1. The first result concerns a weakened version of a Cheeger-type inequality, which does not involve the graph's maximum vertex degree.

LEMMA A.1. *Let $\mathcal{C}$ be a connected finite graph with $|\mathcal{C}| \ge 2$ vertices. Then the smallest nonzero Laplacian eigenvalue $E_{\min}(\mathcal{C})$ is bounded from below according to*

$$(A.1) \qquad E_{\min}(\mathcal{C}) \ge \frac{1}{|\mathcal{C}|^2}.$$

PROOF. The proof is inspired by that of Lemma 1.9 in [10]. It does not involve the maximum vertex degree, though.

Set $n := |\mathcal{C}|$ and let $\mathcal{V} := \{1, \ldots, n\}$ be the vertex set and $\mathcal{E}$ the edge set of $\mathcal{C}$. Elements of $\mathcal{E}$ are denoted by unordered pairs $[i, j]$ of the vertices $i, j \in \mathcal{V}$ they join. The minmax-principle and (2.4) imply that

$$(A.2) \qquad E_{\min}(\mathcal{C}) = \inf_{\varphi \in \mathbb{R}^n \,:\, \sum_{i \in \mathcal{V}} \varphi_i = 0} \frac{\sum_{[i,j] \in \mathcal{E}} (\varphi_i - \varphi_j)^2}{\sum_{i \in \mathcal{V}} \varphi_i^2},$$

where the infimum is taken over $\mathbb{R}^n$ only (instead of $\mathbb{C}^n$), because all eigenvectors of $\Delta(\mathcal{C})$ can be chosen to be real. The other constraint expresses the fact that the nondegenerate zero eigenvalue of the connected graph corresponds to an eigenvector with constant components.

Now, for any given $\varphi \in \mathbb{R}^n$, obeying the orthogonality constraint $\sum_{i \in \mathcal{V}} \varphi_i = 0$, let $u \in \mathcal{V}$ be such that $|\varphi_u| = \max_{i \in \mathcal{V}} |\varphi_i|$. Due to the constraint, there exists $v \in \mathcal{V}$ such that

$$(A.3) \qquad \varphi_u \varphi_v < 0.$$

Let $\mathcal{P}_{\varphi} \subseteq \mathcal{E}$ be the shortest path in $\mathcal{C}$ connecting the vertices $u$ and $v$. Then we have

$$(A.4) \qquad E_{\min}(\mathcal{C}) \geq \inf_{\varphi \in \mathbb{R}^n : \sum_{i \in \mathcal{V}} \varphi_i = 0} \frac{\sum_{[i,j] \in \mathcal{P}_{\varphi}} (\varphi_i - \varphi_j)^2}{|\mathcal{C}| \varphi_u^2}.$$

The triangle and the Cauchy–Schwarz inequality supply us with the estimate

$$(A.5) \quad |\varphi_u - \varphi_v| \leq \sum_{[i,j] \in \mathcal{P}_{\varphi}} |\varphi_i - \varphi_j| \leq \left\{ \sum_{[i,j] \in \mathcal{P}_{\varphi}} (\varphi_i - \varphi_j)^2 \right\}^{1/2} |\mathcal{P}_{\varphi}|^{1/2},$$

where $|\mathcal{P}_{\varphi}|$ stands for the number of edges in $\mathcal{P}_{\varphi}$. Inserting (A.5) into (A.4) and noting $|\mathcal{P}_{\varphi}| < |\mathcal{C}|$, we arrive at

$$(A.6) \qquad E_{\min}(\mathcal{C}) \geq \frac{1}{|\mathcal{C}|^2} \inf_{\varphi \in \mathbb{R}^n : \sum_{i \in \mathcal{V}} \varphi_i = 0} \frac{\varphi_u^2 + \varphi_v^2 - 2\varphi_u \varphi_v}{\varphi_u^2}.$$

The claim now follows from (A.3).    □

The second auxiliary result summarizes the existence, normalization and decay of the cluster-size distribution of Erdős–Rényi random graphs in the subcritical regime.

LEMMA A.2.    *Assume the subcritical regime $p \in ]0, 1[$ and let $\mathcal{C}(1)$ be the maximally connected subgraph of $\mathcal{G}^{(N)}$ containing vertex number one. Then the* cluster-size distribution

$$(A.7) \qquad \tau_n(p) := n^{-1} \lim_{N \to \infty} \mathbb{P}_p^{(N)} \{|\mathcal{C}(1)| = n\} = \frac{1}{n!} n^{n-2} p^{n-1} e^{-np}$$

*exists for every $n \in \mathbb{N}$ and equals the mean number density of tree clusters with $n$ vertices. It is normalized according to*

$$(A.8) \qquad \qquad \sum_{n=1}^{\infty} n \tau_n(p) = 1$$

*and has an exponentially small tail*

$$(A.9) \qquad \qquad \tau_n(p) \leq \frac{1}{\sqrt{2\pi p}} \frac{1}{n^{5/2}} e^{-nf(p)},$$

*with the decay parameter $f(p) := p - 1 - \ln p > 0$.*

PROOF.    The lemma follows from collecting some well-known properties of Erdős–Rényi random graphs in [4]. Fix $n \in \mathbb{N}$ and let $\mathcal{C}(1) \in \mathcal{T}$, respectively $\mathcal{C}(1) \in \mathcal{T}_n$, denote the event that $\mathcal{C}(1)$ is a tree cluster, respectively, a tree cluster with $n$ vertices. Then we have

$$(A.10) \qquad \begin{aligned} \mathbb{P}_p^{(N)} \{|\mathcal{C}(1)| = n\} &= \mathbb{P}_p^{(N)} \{\mathcal{C}(1) \in \mathcal{T}_n\} \\ &\quad + \mathbb{P}_p^{(N)} \{|\mathcal{C}(1)| = n \text{ and } \mathcal{C}(1) \notin \mathcal{T}\}. \end{aligned}$$

Introducing $\chi_A$, the characteristic function of an event $A$, we get an upper bound for the last probability in (A.10):

$$
\begin{aligned}
(A.11) \quad & \mathbb{P}_p^{(N)}\{|\mathcal{C}(1)| = n \text{ and } \mathcal{C}(1) \notin \mathcal{T}\} \\
& \leq \mathbb{P}_p^{(N)}\{\mathcal{C}(1) \notin \mathcal{T}\} \\
& = 1 - \mathbb{P}_p^{(N)}\{\mathcal{C}(1) \in \mathcal{T}\} = 1 - \frac{1}{N} \sum_{j=1}^{N} \mathbb{E}_p^{(N)}\{\chi_{\mathcal{T}}(\mathcal{C}(j))\}.
\end{aligned}
$$

The sum in the last line of (A.11) represents the mean number of vertices on tree clusters in $\mathcal{G}^{(N)}$. Hence, Theorem 5.7(ii) in [4] implies that

$$
(A.12) \qquad \lim_{N \to \infty} \mathbb{P}_p^{(N)}\{|\mathcal{C}(1)| = n \text{ and } \mathcal{C}(1) \notin \mathcal{T}\} = 0
$$

for all $p \in ]0, 1[$. On the other hand, recalling the notation for enumerating clusters above (4.3), we deduce from the equality

$$
\begin{aligned}
(A.13) \quad & \frac{1}{n} \mathbb{P}_p^{(N)}\{\mathcal{C}(1) \in \mathcal{T}_n\} = \frac{1}{Nn} \sum_{j=1}^{N} \mathbb{E}_p^{(N)}\{\chi_{\mathcal{T}_n}(\mathcal{C}(j))\} \\
& = \frac{1}{N} \sum_{k=1}^{K} \mathbb{E}_p^{(N)}\{\chi_{\mathcal{T}_n}(\mathcal{C}_k^{(N)})\}
\end{aligned}
$$

that $n^{-1}\mathbb{P}_p^{(N)}\{\mathcal{C}(1) \in \mathcal{T}_n\}$ equals the mean number density of tree clusters with $n$ vertices in $\mathcal{G}^{(N)}$. Accordingly, (5.1) in [4] yields

$$
\begin{aligned}
(A.14) \quad & \frac{1}{n} \mathbb{P}_p^{(N)}\{\mathcal{C}(1) \in \mathcal{T}_n\} = \frac{1}{N} \binom{N}{n} n^{n-2} \left(\frac{p}{N}\right)^{n-1} \\
& \times \left(1 - \frac{p}{N}\right)^{n(N-n) + \binom{n}{2} - n + 1}.
\end{aligned}
$$

This expression has a limit as $N \to \infty$, which is given by

$$
(A.15) \qquad \tau_n(p) := \frac{1}{n} \lim_{N \to \infty} \mathbb{P}_p^{(N)}\{\mathcal{C}(1) \in \mathcal{T}_n\} = \frac{1}{n!} n^{n-2} p^{n-1} e^{-np}.
$$

Taken together, (A.10), (A.12) and (A.15) establish the first assertion (A.7) of the lemma. The second assertion, the normalization (A.8) follows from (5.6) in [4] and the definition in the equation above (5.5) in [4]. Finally, to prove the decay (A.9) of $\tau_n(p)$, we apply the Stirling inequality $n! \geq (n/e)^n \sqrt{2\pi n} \times \exp\{1/(12n + 1)\}$, see, for example, (1.4) in [4], to (A.15).  $\square$

**Acknowledgment.** Oleksiy Khorunzhiy is grateful to the Ruhr–Universität Bochum for the kind hospitality in June 2004.

O. KHORUNZHIY
UNIVERSITÉ DE VERSAILLES
 ST-QUENTIN-EN-YVELINES
LAMA, BÂTIMENT FERMAT
45 AVENUE DES ETATS-UNIS
78035 VERSAILLES
FRANCE
E-MAIL: khorunjy@math.uvsq.fr

W. KIRSCH
FAKULTÄT UND INSTITUT FÜR MATHEMATIK
RUHR-UNIVERSITÄT BOCHUM
UNIVERSITÄTSSTRASSE 150
44780 BOCHUM
GERMANY
E-MAIL: werner.kirsch@rub.de

P. MÜLLER
INSTITUT FÜR THEORETISCHE PHYSIK
GEORG-AUGUST-UNIVERSITÄT
37077 GÖTTINGEN
GERMANY
E-MAIL: peter.mueller@physik.uni-goettingen.de